\documentclass[aps,pre,twocolumn,showpacs,superscriptaddress,
amsmath,amssymb,floatfix]{revtex4-1}
\pdfoutput=1

\usepackage{epsfig}
\usepackage{graphicx}
\usepackage{epstopdf}
\usepackage{bm}

\usepackage{hyperref}
\usepackage[all]{hypcap}

\begin{document}

\title{How turbulence regulates biodiversity in 
systems with cyclic competition}

\author{Daniel~Gro\v selj}
\affiliation{Max-Planck-Institut f\" ur Plasmaphysik, Boltzmannstra\ss e 2, 
D-85748 Garching, Germany}
\author{Frank~Jenko}%
\affiliation{Max-Planck-Institut f\" ur Plasmaphysik, Boltzmannstra\ss e 2, 
D-85748 Garching, Germany}
\affiliation{Department of Physics and Astronomy, University of California, 
Los Angeles, California 90095-1547, USA}
\author{Erwin~Frey}
\affiliation{Arnold Sommerfeld Center for Theoretical Physics 
and Center for NanoScience, Department of Physics, 
Ludwig-Maximilians-Universit\" at M\" unchen, 
Theresienstra\ss e 37, D-80333 M\" unchen, Germany}

\date{\today}

\begin{abstract}
Cyclic, nonhierarchical interactions among biological species represent a
general mechanism by which ecosystems are able to maintain high levels of 
biodiversity. However, species coexistence is often possible only in 
spatially extended systems with a limited range of dispersal, 
whereas in well-mixed environments models for cyclic competition 
often lead to a loss of biodiversity. Here we consider the dispersal of 
biological species in a fluid environment, where mixing is achieved by a 
combination of advection and diffusion. In particular, we perform a 
detailed numerical analysis of a model composed of turbulent advection, 
diffusive transport, and cyclic interactions among biological species 
in two spatial dimensions and discuss the circumstances under which 
biodiversity is maintained when external environmental conditions, 
such as resource supply, are uniform in space. Cyclic interactions 
are represented by a model with three competitors, resembling the 
children's game of rock-paper-scissors, whereas the flow field is 
obtained from a direct numerical simulation of two-dimensional turbulence 
with hyperviscosity. It is shown that the space-averaged dynamics 
undergoes bifurcations as the relative strengths of advection and diffusion
compared to biological interactions are varied.
\end{abstract}






\maketitle

\section{Introduction}

Several studies have shown that biodiversity in spatially extended population
models can sometimes be maintained even if only a single species is able to 
survive in the well-mixed system \cite{Durrett97, Durrett98, Bracco2000, 
Frean2001, Kerr2002, Karolyi2005, Reichenbach2007, Wang2011}. This is 
typically the case if the interactions among individuals are sufficiently 
local and the system does not display a clear competition hierarchy as in 
the case of cyclic interactions. Cyclic competitions have been frequently 
investigated in discrete lattice models with nearest-neighbor interactions, 
where the dispersal of individuals is local \cite{Tainaka1989, 
Frachebourg1996, Durrett97, Durrett98, Frean2001, Reichenbach2007, 
Perc2007, Reichenbach2008, He2010, Juul2012, Roman2012, 
Avelino2012, Szolnoki2014}. To demonstrate the approach to the well-mixed 
limit in these models one can either increase the effective range of 
interactions towards the total domain size \cite{Durrett98, Frean2001} or 
explicitly consider individuals' mobility by allowing random exchange events 
among adjacent sites, which leads to diffusive transport in the continuum 
limit \cite{Reichenbach2007, Reichenbach2008, He2010, Avelino2012}.
However, a significant part of life on Earth is represented by microorganisms 
dwelling in moving fluid environments, such as the Earth's oceans, where the 
dominant mechanism of transport is typically attributed to fluid turbulence 
\cite{Martin2003125, Pasquero2004, Levy2008}. The latter mechanism of 
microorganisms' transport was a source of inspiration for the work 
presented herein since, despite the fact that cyclic interactions are 
used as a paradigm to explain biodiversity, to our knowledge, only one 
previous study has considered spatial games with cyclic dominance in 
a fluid environment with mixing \cite{Karolyi2005}, and none was devoted 
to the study of cyclic competitions in a turbulent flow so far. 
One of the most intriguing examples of biological communities without a 
clear competition hierarchy inhabiting turbulent aquatic environments is 
represented by marine phytoplankton species that typically compete for a 
limited number of natural resources \cite{Huisman2001a, Barton2010}. 
Apart from marine ecosystems there also seems to be a growing interest 
for biodiversity in the atmosphere, although it remains an open question 
as to whether organisms found in the atmosphere can be regarded as an active 
ecosystem or not~\cite{Burrows2009, Womack2010, DeLeon-Rodriguez2013}.

Fluid motion can have a profound effect on the time evolution of passively 
advected biological populations or chemical substances \cite{Epstein1994, 
Abraham2000, Bracco2000, Koszalka2007, Neufeld2010}. In particular, stirring 
by a fluid flow can lead to transitions between different dynamical regimes 
of reactive systems, mathematically described by the so-called 
reaction-diffusion-advection (RDA) equations \cite{Neufeld2002, Neufeld2003, 
Paoletti2006, Neufeld2012}. From an ecological point of view, these 
transitions correspond to changes in relative species abundance and are 
therefore of crucial importance for studies of biodiversity in moving fluid 
environments. Here we demonstrate that, upon changing the relative strengths 
of advection and diffusion compared to biological interactions, a spatially 
extended population model with cyclic dominance experiences dramatic changes 
in its spatiotemporal dynamics. Distinct dynamical regimes of the system 
are manifested by a rich variety of phenomena such as rotating spiral waves, 
the emergence of periodic oscillations in relative species abundance, and 
transitions into absorbing states where only one species survives. 

To shed some light on cyclic competitions in turbulent flows we adopt a 
minimal biological model with three competitors dominating each other in an 
analogous fashion as in the children's game of rock-paper-scissors, where 
rock crushes scissors, scissors cut paper, and paper covers rock. Known 
examples where cyclic dominance has been identified in interactions 
between three competitors include the mating strategies of lizards in 
the inner Coast Range of California \cite{Sinervo1996}, competitions 
between mutant strains of yeast \cite{Paquin1983}, and bacterial strains 
of toxin-producing \emph{Escherichia coli} 
\cite{Kerr2002, Kirkup2004, Weber2014}. In the spatially extended population 
model, the biological reactions are supplemented with diffusion terms and 
advection by a two-dimensional (2D) turbulent flow. Diffusion can be used 
either to represent random Brownian motion of the advected species or as 
a parametrization for turbulent transport below the model's resolution, 
i.e.~at the scales where three-dimensional fluid motions become 
important~\cite{Martin2003125, Bracco2009}. Diffusion, however small it may 
be, plays in fact a crucial role in the population dynamics of passively 
advected species because it enables particles in nearby fluid 
elements to come in contact and interact \cite{Richards2006, Neufeld2010}. 
In cases where particle inertial effects or microorganism 
motility are considered important, the species may additionally come in 
contact due to an effective compressibility of the flow 
field \cite{Perlekar2010, Benzi2012, Durham2013}. In this study, 
however, these effects are neglected and the tracer velocity field 
is assumed to be incompressible.

It has been previously shown that cyclic interactions in combination with 
diffusion lead to self-organization of the three competitors into rotating 
spiral waves \cite{Reichenbach2007, Reichenbach2008}. Turbulent advection in 
2D flows is, on the other hand, known to produce sharp, patchy distributions 
of biological tracers \cite{Abraham1998, Bracco2009, 
McKiver2009, McKiver2011}. Advective and diffusive transport in systems 
with cyclic competition are therefore drawn towards a complex interplay of 
diverse factors, exhibiting elements of competition as well as of 
cooperation. On one hand, these two processes work together to enhance the 
overall mixing rate, while, on the other hand, they represent two opposing 
mechanisms, favoring either sharp gradients in the subpopulation densities 
or smooth density profiles propagating in the form of traveling waves. 
Various relative strengths of turbulence compared to diffusion can be also 
viewed---in a more loose sense---as different compromises between the random 
Brownian and collective motions of individuals. Our model might therefore 
provide some general insight into situations where the motions of individuals
are adequately described by a single, spatially correlated velocity field in 
the continuum limit. Apart from plankton species in the ocean, an 
interesting example belonging to this general class of biological systems 
is represented by swimming bacteria in dense suspensions \cite{Ishikawa2009}
where the velocity correlation length was shown to depend mainly on the 
bacterial concentration~\cite{Gachelin2014}. We also note that the 
applications of this study concerning pattern formation in fluid flows 
are not limited to biological systems since qualitatively 
similar patterns to the ones observed in our model can also be 
reproduced with the celebrated Belousov-Zhabotinsky 
reaction~\cite{Cross1993, Nugent2004, Kameke2013}. 

In the ocean, advection is recognized as the dominant source of transport, 
and the biological interactions among planktonic organisms typically occur 
on similar time scales as horizontal mixing \cite{Abraham1998, Bracco2009}.
Nevertheless, in order to obtain a thorough understanding of the various 
physical processes involved, and due to possible applications of our work 
outside the field of marine ecology, we perform a comprehensive numerical 
analysis of our model over a wide range of relative advection as well as 
diffusion strengths compared to the biological interactions.  All the 
available simulation results are then used to construct a rough 
nonequilibrium phase diagram of the spatiotemporal dynamics.

The remainder of this article is structured as follows. 
In~Sec.~\ref{sec:model} we present a set of rate equations describing cyclic
interspecies interactions and give details regarding our 2D turbulence 
simulation, before discussing the full set of RDA equations used to model 
the dynamics of the spatially extended system. In~Sec.~\ref{sec:results} 
we perform a detailed analysis of the numerical results. The complex 
spatiotemporal patterns are first inspected through snapshots of the 
solutions and by means of space-time autocorrelation functions. 
Afterwards, adopting a recently introduced method developed in the 
context of interacting particle systems \cite{Rulands2013}, we show 
that the most pronounced \emph{qualitative} changes in the system's 
spatiotemporal dynamics correspond to bifurcations of 
the space-averaged dynamics. The main conclusions 
are given in~Sec.~\ref{sec:conclusions}.

\section{\label{sec:model} Model}

\subsection{Rate equations}

We study a biological population comprised of three species $A$, $B$, and $C$
that cyclically dominate each other. In addition, the individuals from each 
subpopulation are able to reproduce if an empty spot $\varnothing$ is 
available. The complete model composed of selection and reproduction 
processes is described by the following reaction 
scheme~\cite{Reichenbach2008, Szolnoki2014}:
\begin{align}
&& AB\stackrel{\sigma}{\longrightarrow}A\varnothing, && 
A\varnothing\stackrel{\mu}{\longrightarrow}AA,\nonumber&&\\
&& BC\stackrel{\sigma}{\longrightarrow}B\varnothing, && 
B\varnothing\stackrel{\mu}{\longrightarrow}BB, &&\label{eq:reac_rps}\\
&& CA\stackrel{\sigma}{\longrightarrow}C\varnothing, && 
C\varnothing\stackrel{\mu}{\longrightarrow}CC, \nonumber&&
\end{align}
where $\sigma$ and $\mu$ are the selection and reproduction rates, 
respectively. If the size of each subpopulation is macroscopically large, 
such that the relative fluctuations arising from stochastic effects are 
small, the system's time evolution may be treated as deterministic, and 
the discrete distributions of individuals belonging to each species can 
be replaced by the mean densities $a$, $b$, and $c$ of subpopulations 
$A$, $B$, and $C$, respectively. In the well-mixed limit, the mean 
population densities are governed by the following set of rate equations:
\begin{align}
\partial_t a & = \mu a(1 - \rho) - \sigma ac,\nonumber\\
\partial_t b & = \mu b(1 - \rho) - \sigma ba,\label{eq:rps}\\
\partial_t c & = \mu c(1 - \rho) - \sigma cb,\nonumber
\end{align}
where $\rho=a+b+c$ represents the total population density.

\begin{figure}[ht]
\includegraphics{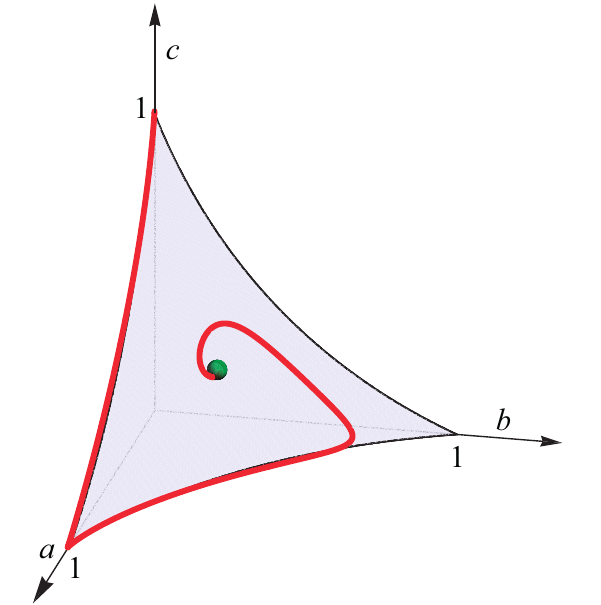}
\caption{\label{fig:invManifold}(Color online) Evolution of subpopulation 
densities $(a,b,c)$ on the invariant manifold for equal selection and 
reproduction rates ($\mu=\sigma$). The red (gray) curve shows a trajectory 
starting in the vicinity of the reactive fixed point.}
\end{figure}

Equation \eqref{eq:rps} is a special case of a three-species population 
model first studied by May and Leonard \cite{May1975}. For the above system, 
May and Leonard reported four nontrivial fixed points \cite{May1975}: three 
single-species equilibrium points, $(1,0,0)$, $(0,1,0)$, $(0,0,1)$, and an 
unstable reactive fixed point, $\frac{\mu}{3\mu+\sigma}(1,1,1)$, where all 
three subpopulations coexist. The single-species fixed points correspond to 
absorbing states that can never be left by the system's dynamics once they 
are reached. Solutions of Eq.~\eqref{eq:rps} starting in the vicinity of the
coexistence point form so-called \emph{heteroclinic orbits} 
\cite{May1975, Hofbauer1998}: the trajectories $(a(t),b(t),c(t))$ spiral 
outwards from the coexistence point and come ever closer to the 
single-species equilibrium points but never converge to any of them. 
However, the time spent in the vicinity of the single-species fixed 
points increases proportionally with time. In consequence, one of the 
species will sooner or later dominate the other two over times which 
are much longer than any biological time scale of interest, even though 
there can be no winner in the strict limit $t\!\to\!\infty$. It should 
also be noted that any inclusion of (demographic) noise in the deterministic 
model \eqref{eq:rps} will inevitably lead to the complete extinction of all 
but one species in a finite time \cite{Frey2010}. \mbox{Reichenbach} 
et al.~\cite{Reichenbach2008} have investigated system \eqref{eq:rps} 
further and showed that all solutions of Eq.~\eqref{eq:rps} decay onto 
a 2D invariant manifold (a subspace left invariant by the system's 
time evolution), which contains all four nontrivial fixed points. The 
invariant manifold together with an example of a phase space trajectory 
starting in the vicinity of the reactive fixed point is 
shown in Fig.~\ref{fig:invManifold}.

\subsection{Turbulence model}

The turbulent velocity field is obtained from solutions 
of the incompressible, 2D Navier-Stokes equations with hyperviscosity and 
large-scale, random forcing. Real flows in nature are always, at least to 
some extent, three dimensional. However, the particular type of flow was 
primarily chosen to fit the needs of a model for the stirring of 
microorganisms by large-scale geophysical flows. These flows are strongly 
anisotropic due to geometrical constraints (large horizontal scale compared 
to the fluid's depth) and body forces acting on the fluid (Coriolis force, 
density stratification) and may be treated as 2D in the first approximation 
\cite{Salmon1998, Danilov2000}. Fluid turbulence constrained to two spatial 
dimensions is characterized by many unique features such as an inverse 
cascade of kinetic energy to large scales, a cascade of enstrophy to small 
scales, and strong, long-lived vortices comparable to the size of 
the energy injection scale \cite{Kraichnan1967, davidson, Boffetta2012}. 
With the random forcing being concentrated at large scales, the flow 
field obtained from our simulations is smooth and limited to the 
direct 2D turbulence enstrophy cascade. Given the typical 
forcing scales in the ocean ($\sim50$~km) \cite{Bracco2000}, 
large-scale forcing constraints the simulation domain size within the 
ocean mesoscale range, where horizontal advection is aptly described by 
the (standard) 2D Navier-Stokes equations \cite{Bracco2009}. 
We also note that similar modeling approaches have been used in 
many previous studies of population dynamics over large horizontal scales 
in the ocean~\cite{Abraham1998, Bracco2000, Koszalka2007, Bracco2009, 
McKiver2009, McKiver2011}. 

In two dimensions the Navier-Stokes equations are most conveniently 
solved by integrating the vorticity equation~\cite{Boffetta2012}
\begin{equation}
\partial_t\omega + {\bf v}\cdot\nabla\omega = 
{\mathcal D} + \xi,\label{eq:vorticityNum}
\end{equation}
where ${\bf v}=(v_x, v_y)$ is the 2D velocity field, 
$\omega = \partial_x v_y \allowbreak- \partial_y v_x$ is the (scalar) 
vorticity, $\mathcal D$ represents dissipation terms, and $\xi$ is the 
external forcing. For the dissipation, we use a sum of hyperviscosity
and linear friction given by 
${\mathcal D} = \nu\mathop{\Delta^{3}}\!\omega  - \alpha\omega$. The viscous 
term of the Navier-Stokes equations is frequently replaced by higher 
powers of the Laplacian in turbulence simulations because it is possible to 
achieve higher effective Reynolds numbers at a given spatial resolution in 
this way \cite{Lindborg2000, Bracco2004}. The second source of 
dissipation---the linear drag term---provides a large-scale energy sink, 
necessary to reach a statistically steady state in 2D turbulence simulations 
due to the flow of energy to large scales~\cite{davidson, Boffetta2012}. 

Equation \eqref{eq:vorticityNum} is solved with a pseudospectral method on a 
doubly periodic square domain at resolution $768^2$ using the exponential 
time differencing fourth-order Runge-Kutta scheme \cite{Kassam2005} for 
the time integration. The random forcing is applied in spectral space by 
adapting a general forcing scheme for three-dimensional turbulence, 
introduced by \mbox{Alvelius} \cite{Alvelius1999}, to the 2D case. The 
forcing power spectrum is restricted to a narrow range of wave numbers with 
a peak at $k_{\mathrm f} = 2\pi/\ell_{\mathrm f}$, where $l_{\mathrm f}$ is a
characteristic forcing length scale. Following the approach of 
Ref.~\cite{Alvelius1999}, the time and length units of the simulation are 
fixed by the choice $\ell_{\mathrm f} \equiv 1$ and $P \equiv 1$, where 
$P$ is the average external power input.

\begin{figure}[ht]
\includegraphics{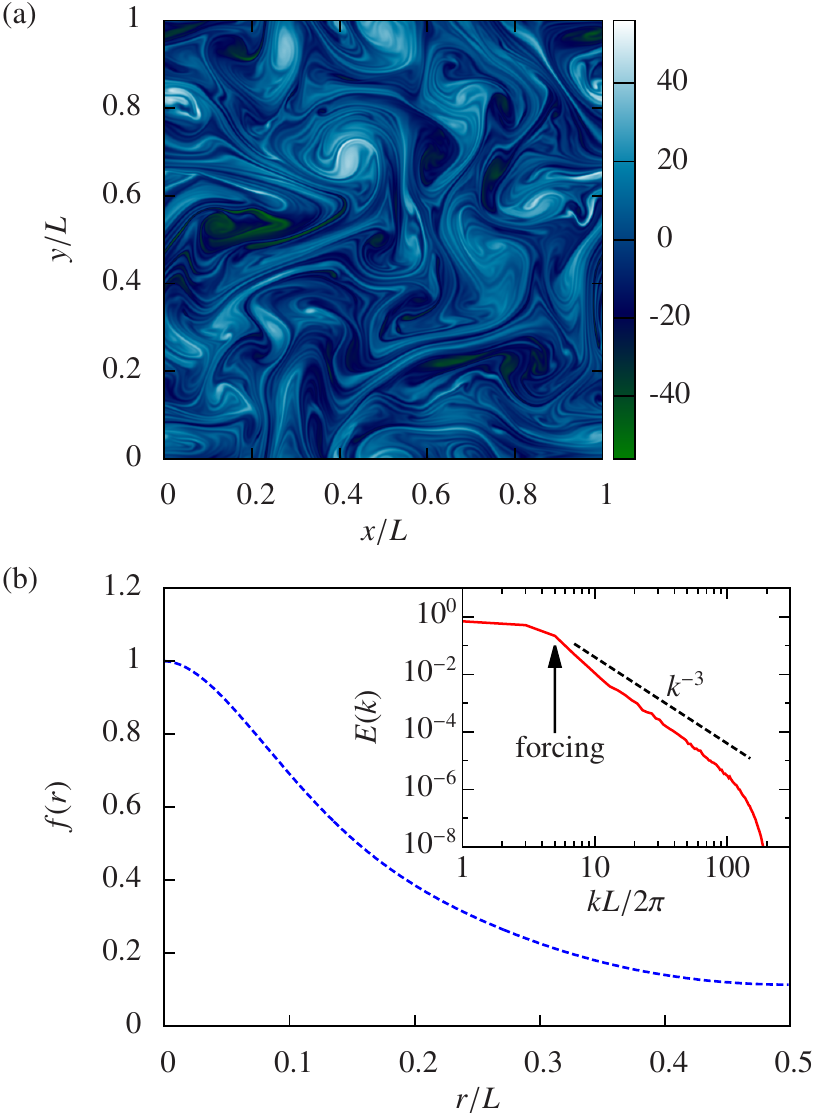}
\caption{\label{fig:turbulencePic}(Color online) Results of the 2D turbulence 
simulation. (a) Snapshot of the vorticity field in a statistically steady 
turbulent state. (b) Longitudinal velocity correlation function. The inset 
shows the turbulence energy spectrum.}
\end{figure}

In the chosen units, the simulation domain edge length, hyperviscosity, and 
drag coefficient were set to $L=5$, $\nu=3\times10^{-13}$, and $\alpha=0.13$, 
respectively. Starting from an initial zero vorticity, 
Eq.~\eqref{eq:vorticityNum} was integrated until a statistically steady 
state, characterized by a steady value of the total kinetic energy, was 
reached. The generated vorticity profile 
(Fig.~\hyperref[fig:turbulencePic]{\ref*{fig:turbulencePic}(a)}) was then 
used as the initial condition for vorticity in the simulations of cyclic 
competitions in a turbulent flow (see~Sec.~\ref{ssec:rpsTurbModel}). To 
confirm that our numerical solution is consistent with well-known results 
from the literature, we computed the kinetic energy spectrum $E(k)$ and the 
longitudinal velocity correlation function 
$f(r) = \langle{v_i({\bf r}',t)v_i({\bf r}' + r\hat{{\bf e}}_i,t)\rangle}
/\langle v_i^2\rangle$ 
(Fig.~\hyperref[fig:turbulencePic]{\ref*{fig:turbulencePic}(b)}), where the 
brackets $\langle\cdot\cdot\cdot\rangle$ denote a space-time average, the 
index $i$ represents the $x$ or $y$ direction, and $\hat{\bf e}_i$ is the 
unit vector. We found  $f(r)$ to be a non-negative, monotonically decreasing 
function with a correlation length (the length $\ell_c$ at which 
$f(\ell_c) = \exp(-1)$) close to the forcing length scale. The estimated 
turbulence energy spectrum has a slope close to $-3$ on the logarithmic 
graph, in the wave number range between the energy injection scale and 
hyperviscous dissipation scale. These results are in good agreement with 
theoretical predictions \cite{Kraichnan1967, davidson} and other numerical 
simulations~\cite{Lindborg2000, Boffetta2010}.

\subsection{\label{ssec:rpsTurbModel} The spatially extended population model}

In the spatially extended model, we combine the differential equations 
for species competition \eqref{eq:rps} and the turbulence model 
\eqref{eq:vorticityNum} into a set of RDA equations for the subpopulation 
densities $a({\bf r},t)$, $b({\bf r},t)$, $c({\bf r},t):$
\begin{align}
\partial_t a + {\bf v}\cdot\nabla a & = \mu a(1 - \rho) - 
\sigma ac + D\Delta\,\! a,&\nonumber\\
\partial_t b + {\bf v}\cdot\nabla b & = \mu b(1 - \rho) - 
\sigma ba + D\Delta\,\! b,&\label{eq:rpsTurb}\\
\partial_t c + {\bf v}\cdot\nabla c & = \mu c(1 - \rho) - 
\sigma cb + D\Delta\,\! c,&\nonumber
\end{align}
where $D$ is the diffusion constant and the flow field ${\bf v}({\bf r}, t)$ 
is determined at each time instant from Eq.~\eqref{eq:vorticityNum}. 

The corresponding discretized version of system \eqref{eq:rpsTurb} as well as
the continuous model have been extensively studied by \mbox{Reichenbach} et 
al.~\cite{Reichenbach2007, Reichenbach2007a, Reichenbach2008} for the case 
with no fluid flow. The authors of Refs.~\cite{Reichenbach2007, 
Reichenbach2007a, Reichenbach2008} have shown that mobile individuals 
exhibiting cyclic dominance are able to coexist up to some critical value of 
species mobility (characterized by an effective diffusion constant in the 
continuum limit). Within the coexistence phase the fields $a({\bf r},t)$, 
$b({\bf r},t)$, and $c({\bf r},t)$ self-organize---in two dimensions---into 
rotating spiral waves. As the species mobility increases, the spirals' 
wavelength grows proportionally to $\sqrt{D}$ until the patterns outgrow the 
system size. The state in which spirals are absent corresponds to the 
well-mixed system \eqref{eq:rps} where only one subpopulation survives. 
Taking into account the main features of the reaction-diffusion part of 
Eq.~\eqref{eq:rpsTurb}, the natural time and length scale of the spatially 
extended model appear to be the spirals' rotation period 
$\mathop{T_0}(\mu,\sigma)$ and the linear size of the simulation domain $L$. 
We shall make use of these units in the following analysis of our numerical 
results. It is also worth emphasizing that $T_0$ is uniquely determined by 
$\mu$ and $\sigma$ alone \cite{Reichenbach2008}, even though rotating spirals 
can only emerge in the presence of diffusion.

For a given type of 2D flow and for a fixed ratio $\sigma/\mu$, the solutions
of Eq.~\eqref{eq:rpsTurb} are characterized by two dimensionless parameters, 
which can be constructed by assigning a characteristic time scale to each of 
the three physical phenomena (reactions, diffusion, and advection) and 
comparing these scales to each other. Here we choose to analyze our results 
in terms of the parameters
\begin{align}
&& \mathrm{Da} = \tau_{\mathrm f}/\tau_{\mathrm r} && \text{and} 
&& K_{\mathrm d} = \tau_{\mathrm d}/\tau_{\mathrm r}, &&
\label{eq:paramDef}
\end{align}
where $\tau_{\mathrm r}$, $\tau_{\mathrm d}$, and $\tau_{\mathrm f}$ are a 
characteristic reaction, diffusion, and flow time scale, respectively. The 
ratio $\mathrm{Da}$ is known in literature as the \emph{Damk\" ohler number}
\cite{Neufeld2010}. We adopt a common definition for the flow time scale 
given by $\tau_{\mathrm f} = \ell_{\mathrm f}/u$, where $u$ is the 
root-mean-square velocity of the flow \cite{McKiver2009, Neufeld2010}. 
For $\tau_{\mathrm r}$, we use the definition $\tau_{\mathrm r} =  T_0$ 
because this appears to be the slowest reaction time scale of 
Eq.~\eqref{eq:rpsTurb}, and it is reasonable to expect that the reactions 
will balance turbulent advection only when the slowest reaction time scale 
is able to keep up with the flow. We estimated the spirals' rotation period 
from simulation runs performed with ${\bf v}({\bf r},t)= 0$ and 
found $T_0\approx 61.5/\mu$ for $\mu=\sigma$, which is in good agreement 
with \mbox{Reichenbach} et al.~\cite{Reichenbach2008}. The diffusion time 
scale $\tau_{\mathrm d}$ is defined by the ratio $L^2/D$ in order to 
make $K_{\mathrm d}$ independent of any parameters of the flow. Hence, 
$K_{\mathrm d}$ gives the inverse of the diffusion constant in 
units of $L^2/T_0$.

Equation \eqref{eq:rpsTurb} is solved on a doubly periodic square domain 
using a second-order operator splitting (Strang splitting) approach 
\cite{Hundsdorfer2003} which treats separately the reaction and 
advection-diffusion part of Eq.~\eqref{eq:rpsTurb}. The advection-diffusion 
terms are integrated with a hybrid method, introduced by \mbox{Spiegelman} 
and \mbox{Katz} \cite{Spiegelman2006}, which combines the semi-Lagrangian 
scheme for advection with the Crank-Nicolson algorithm for the diffusion 
equation. In the semi-Lagrangian method, we use a second-order midpoint 
iteration technique for finding the departure point of each fluid parcel 
at previous time step \cite{Staniforth1991}, together with bicubic 
interpolation for approximating the values of the advected fields at 
departure points. The bicubic interpolation is constructed from a 
series of one-dimensional cubic spline interpolations with fourth-order 
central difference estimates of the derivatives at the interpolating 
nodes. To reduce spurious oscillations, which typically arise from 
standard high-order interpolations near sharp gradients of the 
concentration fields, a monotonicity-preserving modification of the 
derivatives is used for each one-dimensional cubic interpolation 
\cite{Fritsch1980}. The reaction terms are integrated independently 
of the advection-diffusion part with a second-order Runge-Kutta scheme. 
In order to make our numerical method consistent with the externally 
supplied time-dependent flow field ${\bf v}({\bf r},t)$, the same 
spatial resolution as in the 2D turbulence simulation 
($768^2$ grid points) is used to solve Eq.~\eqref{eq:rpsTurb}.

In all simulations, we use initial conditions of the form 
\begin{equation}
(a^* + \delta\xi_a,\, b^* + \delta\xi_b,\, c^* + \delta\xi_c),
\label{eq:rpsInit}
\end{equation}
where $(a^*,b^*,c^*)=\frac{\mu}{3\mu+\sigma}(1,1,1)$ is the reactive fixed 
point of Eq.~\eqref{eq:rps}, $\{\xi_s\}$ are randomly distributed numbers 
between $-1$ and $1$, and $\delta$ is the amplitude of fluctuations around 
the reactive fixed point. The random numbers are generated independently for 
each point on the computational grid so that no spatial correlations are 
present in the initial conditions. We also avoid using the same flow time 
evolution for all simulation runs by initializing the random turbulence 
forcing term differently for each run. It is important to note that various 
initial conditions and flow realizations should be considered for studies of 
ecosystem stability because any particular solution of Eq.~\eqref{eq:rpsTurb} 
might show stability properties that---on a given time scale---significantly 
differ from the statistical average over many realizations.

\section{\label{sec:results} Results}

To investigate the characteristics of our spatially extended population model
we performed over 250 simulation runs for the system \eqref{eq:rpsTurb}. In 
most cases, the equations were integrated over a time 
$T\approx 23\tau_{\mathrm r}$. In order to determine the circumstances under 
which the spatial degrees of freedom facilitate a significant improvement of 
ecosystem stability, it is sufficient to consider time scales which are only 
about an order of magnitude larger than $\tau_{\mathrm r}$, since the 
heteroclinic orbits of the well-mixed system \eqref{eq:rps} typically require
less than $\tau_{\mathrm r}$ to reach the boundaries of the phase space. In 
the following, we refer to the time scales which are only about an order, or 
perhaps a few orders, of magnitude larger than $\tau_{\mathrm r}$ as 
to the biological or ecological \emph{time scales of interest}. Moreover, 
the term ``long-time regime'' should be in the following understood only 
in the context of such time scales.

In our simulations, the amplitude of fluctuations around the reactive fixed 
point in the initial conditions was set to $\delta=2.5\times10^{-3}$, 
and we have always used equal selection and reproduction rates ($\mu=\sigma$).
\mbox{Reichenbach} et al.~\cite{Reichenbach2008} have shown that different 
choices of the ratio $\sigma/\mu$ do not qualitatively change the system's 
dynamics in the limit $\mathrm{Da}\!\to\!\infty$. It seems reasonable to 
expect this to be true in general for all Damk\" ohler numbers, although 
we have not explicitly considered various choices of $\sigma/\mu$ in our 
simulations. Different values of $\mathrm{Da}$ for fixed $K_{\mathrm d}$ 
were in the simulations achieved by rescaling the entire right-hand 
side of Eq.~\eqref{eq:rpsTurb} while keeping the parameters of the 
flow unchanged. The same effect could have been alternatively achieved 
by varying the magnitude of the flow field while keeping the parameters 
of the reaction-diffusion part fixed. 

Different aspects of the solutions are presented in two subsections. In 
Sec.~\ref{ssec:spatioTemporal} we describe the spatiotemporal dynamics for 
those choices of $\mathrm{Da}$ and $K_{\mathrm d}$ that give rise to a 
heterogeneous spatial structure (i.e.~the system is able to maintain a state 
of biodiversity). In the following, we call the latter range of values for 
$\mathrm{Da}$ and $K_{\mathrm d}$ the \emph{species coexistence region}. A 
more formal definition of this term is given later in 
Sec.~\ref{ssec:ecosystemStability}, where we discuss the transitions between 
different dynamical regimes and present a rough nonequilibrium phase 
diagram of the spatiotemporal dynamics.

\subsection{\label{ssec:spatioTemporal} System's spatiotemporal evolution}

\begin{figure}[ht]
\includegraphics{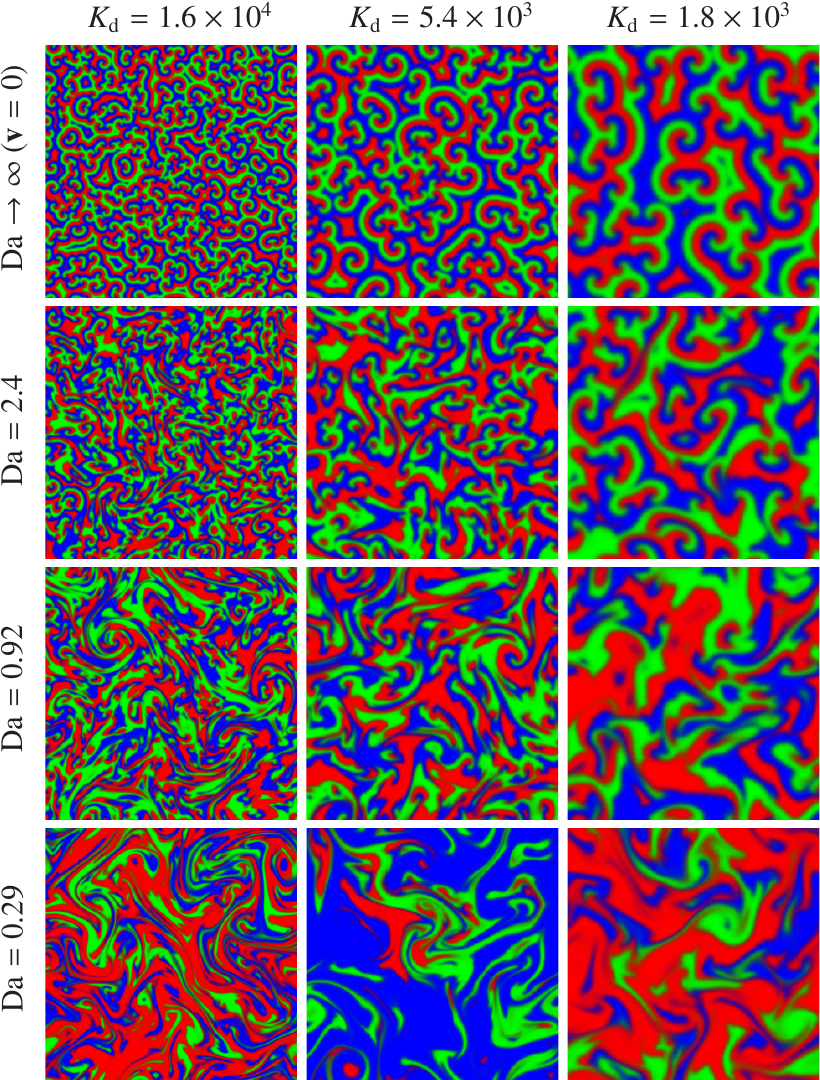}
\caption{\label{fig:RGBplots}(Color online) Snapshots of the concentration 
fields $a$, $b$, and $c$ for different Damk\" ohler numbers 
$\mathrm{Da}$ and ratios of the diffusion to reaction time scale 
$K_{\mathrm d}$. Each subpopulation density is represented by 
its own color channel (gray tone)---red (medium gray) for $a$, 
green (light gray) for $b$, and blue (dark gray) for $c$.}
\end{figure}

The spatial structure of solutions can be qualitatively explored through 
snapshots of the subpopulation densities $a({\bf r}, t)$, $b({\bf r},t)$, 
and $c({\bf r},t)$. Figure \ref{fig:RGBplots} shows a selection of long-time 
snapshots of the solutions. In the long-time regime, the three 
species occupy separate parts of the (periodic) domain, forming various types
of patterns which give qualitative insight into the underlying character of 
spatial transport. In the top row of Fig.~\ref{fig:RGBplots}, we show 
solutions obtained in the absence of fluid flow in order to draw a clear 
picture of the differences between the full set of RDA equations and the 
reaction-diffusion dynamics with ${\bf v}({\bf r},t)=0$. The patterns 
resulting from the interplay between reactions, diffusion, and turbulent 
advection are in general much different from those induced by diffusive 
transport alone because the chosen flow field is correlated on the largest 
scale resolved by the simulations. For moderately large Damk\" ohler numbers 
(Fig.~\ref{fig:RGBplots}, second row), one can observe a collection of 
irregular spiral shapes which are rendered unstable by the stretching and 
folding of material lines in 2D turbulence (see also Supplemental Material 
\footnote{See Supplemental Material at 
[URL will be inserted by publisher] for movies of the system's 
spatiotemporal evolution.}, Movie 1). In other words, these spirals have a 
finite lifetime. However, new spirals are spontaneously formed at a similar 
rate as the old ones are being destroyed. For Damk\" ohler numbers around 
$\mathrm{Da}\approx 1$ (Fig.~\ref{fig:RGBplots}, third row), the 
reaction-diffusion dynamics and turbulent advection are found to be in an 
approximate dynamic balance (Supplemental Material 
\cite{Note1}, Movie 2). This statement will be clarified 
later when we examine the time autocorrelation functions. As the 
Damk\" ohler number is decreased even further (Fig.~\ref{fig:RGBplots}, 
bottom row), fluid mixing significantly increases the effective range 
of species' interactions, leading to collective oscillations in relative 
species abundance on the largest scales (Supplemental Material 
\cite{Note1}, Movie 3). This phenomenon will be discussed in 
more detail later on.

\begin{figure}[ht]
\includegraphics{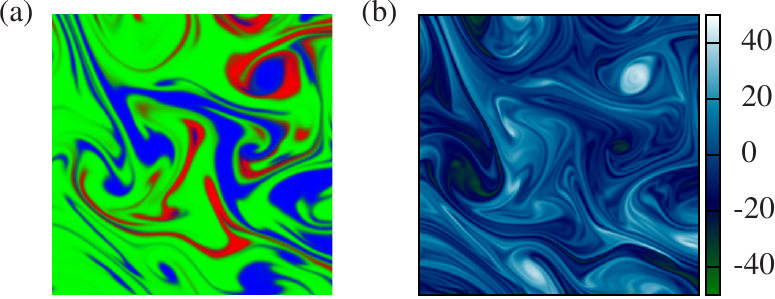}
\caption{\label{fig:RGBAndVorticity}(Color online) Adaptation of 
spatial patterns to the vorticity field structure in the regime 
of long reaction-diffusion time scales compared to the characteristic
flow time scale. (a) Snapshot of the subpopulation densities $a$, $b$, and $c$.
Different colors (gray tones) should be interpreted in the same way as in 
Fig.~\ref{fig:RGBplots}. (b) The vorticity field of 2D turbulence. 
Only 1/4 of the whole domain is shown in (a) and (b). Both snapshots 
are taken at the same time of the simulation run.}
\end{figure}

The influence of turbulent transport on the system's dynamics is strongest 
for low $\mathrm{Da}$ and high $K_{\mathrm d}$. In this regime, the 
subpopulation densities are expected to behave effectively as passive 
(weakly diffusive) tracers on the time scale of the flow $\tau_{\mathrm f}$, 
adapting a structure similar to that of the vorticity field 
\cite{Neufeld2010}. Figure~\ref{fig:RGBAndVorticity} compares a solution of 
Eq.~\eqref{eq:rpsTurb} for $\mathrm{Da}=0.18$ and 
$K_{\mathrm d}= 1.6\times10^4$ with the instantaneous vorticity field and 
confirms this prediction. However, on the time scale of the reaction 
$\tau_{\mathrm r}$ the concentration fields are highly sensitive to fluid 
mixing and develop large oscillations in relative species abundance as 
already mentioned above.

\begin{figure}[ht]
\includegraphics{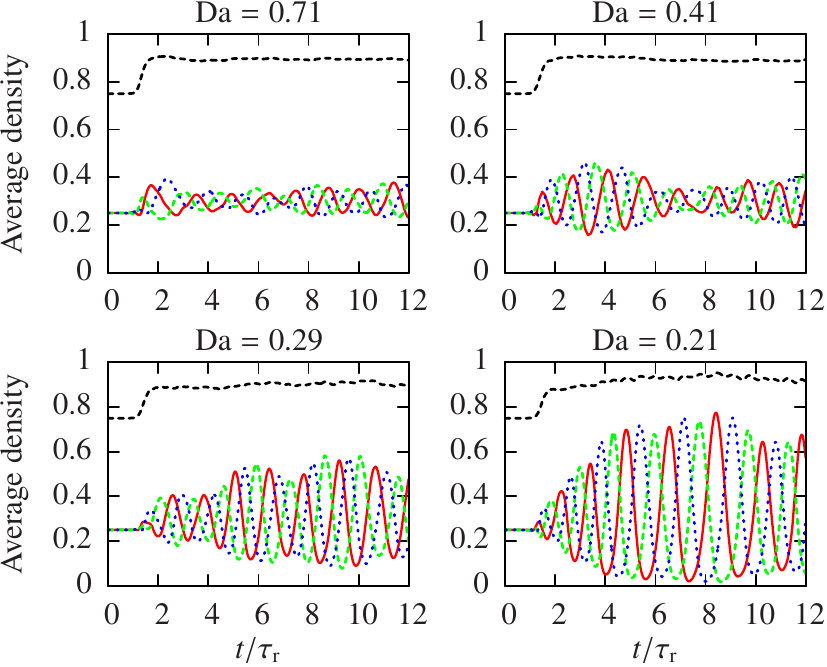}
\caption{\label{fig:avgCon}(Color online) Transition to low 
Damk\" ohler numbers where space-averaged concentrations display large 
periodic oscillations. The solid red (medium gray), dashed green 
(light gray), and dotted blue (dark gray) lines display the 
space-averaged subpopulation densities. The dashed black lines show 
the total space-averaged density. The diffusion to 
reaction time scale ratio $K_{\mathrm d}$ was set to 
$K_{\mathrm d}=8.1\times10^3$ in all cases.}
\end{figure}

The most intriguing questions regarding the system's time evolution are those
related to the long-time maintenance of biodiversity. In the species 
coexistence region, the total space-averaged density $\mathop{\overline\rho}$
always reaches a nearly steady value close to $0.9$. On the contrary, the 
space-averaged subpopulation densities $\mathop{\overline a}$, 
$\mathop{\overline b}$, and $\mathop{\overline c}$ never settle to a steady 
value, but rather oscillate around their space-time average; approximately 
1/3 of the total density. 
The time evolution observed for low Damk\" ohler numbers deserves some 
special attention. In this regime, the space-averaged concentrations display 
surprisingly regular periodic oscillations, and the three subpopulations 
cyclically dominate the total biomass of the system (Fig.~\ref{fig:avgCon}). 
A large global concentration of one of the subpopulations does not 
necessarily lead to a loss of biodiversity because an abundant species 
represents a convenient ``spreading medium'' for its superior competitor 
that can easily outperform the first species and become abundant itself 
before it is in turn replaced by the third species, and so on. However, 
when the relative strength of mixing is increased further, the amplitudes 
of oscillations approach the lower bound of the total space-averaged 
concentration $\mathop{\overline\rho}$, so that the probability of extinction
increases, until species coexistence becomes almost impossible.

Collective oscillations in cyclic competitions with fluid mixing have also 
been reported by \mbox{K\' arolyi} et al.~\cite{Karolyi2005}. The model 
studied in Ref.~\cite{Karolyi2005} was composed of cyclic interactions 
between three species and an analytically prescribed unidirectional shear 
flow with a changing direction. Like in our turbulence model, large-scale 
correlations were present in the velocity field used in 
Ref.~\cite{Karolyi2005}. Interestingly, transitions to states with global 
oscillations have also been observed in studies of cyclic competitions on 
regular small-world networks, where a given portion of randomly chosen 
nearest-neighbor links is replaced with long-range links 
\cite{Szabo2004a, Rulquin2014}. In view of these previous works, our results 
provide further evidence that the collective oscillations are a robust 
phenomenon, unaffected by the details of cyclic interactions, as long as 
there exists a mechanism capable of mediating interactions among spatially 
separated parts of the system. For sufficiently small velocity field 
correlation lengths in statistically stationary and homogeneous flows, 
however, it should be possible to approximate advection with an effective 
diffusion, thereby eliminating the possibility of collective oscillations. 
Instead, the effect of advection in this case would be to increase the size 
of spiral patterns observed in the absence of fluid flow. This phenomenon 
was recently demonstrated in experiments by von \mbox{Kameke} 
et al.~\cite{Kameke2013} with the pattern-forming Belousov-Zhabotinsky 
reaction in a quasi-2D turbulent flow.

To gain a more quantitative understanding of the system's spatiotemporal 
dynamics, we computed the normalized space-time autocorrelation functions 
\begin{align}
\mathop{C_{ss}}(|{\bf r} - {\bf r}'|, t, t') \equiv  
&\tfrac{1}{\sigma^2} \langle s({\bf r},t)s({\bf r}',t')\rangle &\nonumber\\
& - \tfrac{1}{\sigma^2}\langle s({\bf r},t) \rangle
\langle s({\bf r}',t') \rangle, &
\label{eq:autocorr}
\end{align}
where $s\in\{a,b,c\}$, $\sigma^2 = \langle s({\bf r}, t)^2\rangle - 
\langle s({\bf r}, t)\rangle^2$, and the brackets 
$\langle\cdot\cdot\cdot\rangle$ should be in principle understood as 
ensemble averages over all possible realizations of the flow and over all 
initial conditions. When the probability that one (or two) of the species 
will go extinct becomes small on any biologically reasonable time scale, 
the autocorrelations in the system's long-time regime may be approximated 
with finite-time averages. Under such circumstances, the temporal part of 
$\mathop{C_{ss}}(|{\bf r} - {\bf r}'|, t, t')$ in the long-time regime 
depends only on $|t-t'|$. For large Damk\" ohler numbers, the solutions of 
Eq.~\eqref{eq:rpsTurb} strongly depend on initial conditions which raises 
a concern regarding the validity of approximating Eq.~\eqref{eq:autocorr} 
with a time average. Nevertheless, we found that various initial conditions 
in the form of expression \eqref{eq:rpsInit} give very similar estimates for 
$\mathop{C_{ss}}(|{\bf r} - {\bf r}'|, t, t')$, even in the high 
$\mathrm{Da}$ limit. Since the ecosystem model \eqref{eq:rpsTurb} is 
homogeneous, the autocorrelations may also be evaluated with the help of 
space averages. However, space averages alone generally do not give 
sufficiently accurate results due to the presence of various finite-size 
effects in the solutions of Eq.~\eqref{eq:rpsTurb}.

\begin{figure}[ht]
\includegraphics{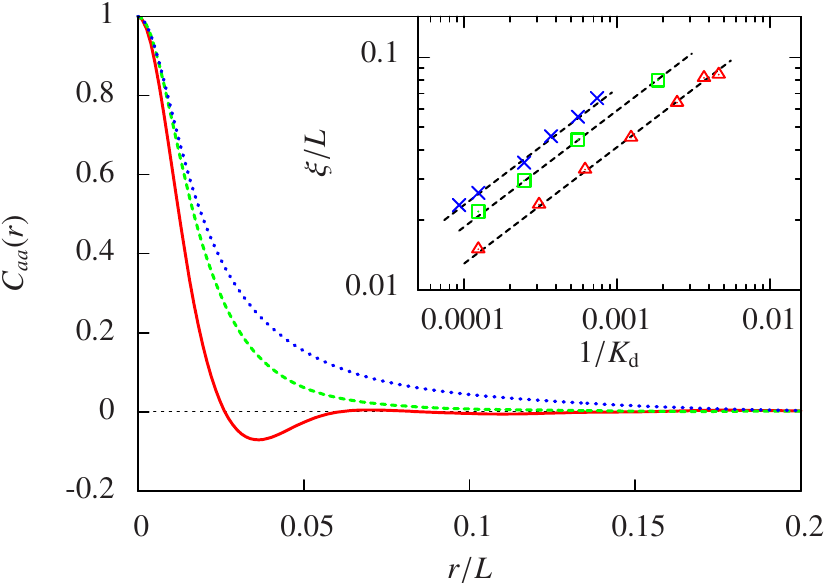}
\caption{\label{fig:rpsTurbLCorr}(Color online) Spatial autocorrelation 
functions for different Damk\" ohler numbers $\mathrm{Da}$. 
The correlations $C_{aa}(r)$ are shown for $\mathrm{Da}=4.7$ 
(solid red line), $\mathrm{Da}=0.92$ (dashed green line), and 
$\mathrm{Da}=0.29$ (dotted blue line). The diffusion to 
reaction time scale ratio $K_{\mathrm d}$ was kept fixed at 
$K_{\mathrm d}=8.1\times10^3$ in all cases. For higher values of 
$\mathrm{Da}$, the autocorrelations suddenly develop a local minimum 
which emerges due to the presence of rotating spiral patterns. 
The inset shows the correlation lengths as functions of 
$1/K_{\mathrm d}$ for $\mathrm{Da}=4.7$ (red triangles), 
$\mathrm{Da}=0.92$ (green squares), and $\mathrm{Da}=0.29$ (blue crosses). 
The dashed lines on the logarithmic graph have a slope of $1/2$ which means 
that all correlation lengths scale as $\xi\sim 1/\sqrt{K_{\mathrm d}}$, 
albeit with a different proportionality factor for each $\mathrm{Da}$.}
\end{figure}

Spatial autocorrelation functions $C_{ss}(|{\bf r} - {\bf r}'|)\equiv 
C_{ss}(|{\bf r} - {\bf r}'|, t, t)$ for different choices of $\mathrm{Da}$ 
are shown in Fig.~\ref{fig:rpsTurbLCorr}. The functions 
$C_{ss}(|{\bf r} - {\bf r}'|)$ were calculated from a space average and an 
additional time average over time $T\approx 13\tau_{\mathrm r}$, once the 
total population density had reached a steady value. In this way, we were 
able to obtain well-behaved estimates of $C_{ss}(|{\bf r} - {\bf r}'|)$ 
with a monotonically decreasing correlation length $\xi$ (the length $\xi$ 
at which $C_{ss}(\xi) = \exp(-1)$) as a function of $\mathrm{Da}$ and 
$K_{\mathrm d}$. The correlation length falls with $\mathrm{Da}$ 
($K_{\mathrm d}$) because an increase of $\mathrm{Da}$ ($K_{\mathrm d}$) 
generally corresponds to a reduced total mixing rate. It has been previously 
shown that $\xi$ scales with the square root of the diffusion constant in 
the absence of fluid flow \cite{Reichenbach2008}. Our results suggest that 
the scaling relation $\xi\sim \sqrt D\sim 1/\sqrt{K_{\mathrm d}}$ remains 
valid in general for any choice of $\mathrm{Da}$.

\begin{figure}[ht]
\includegraphics{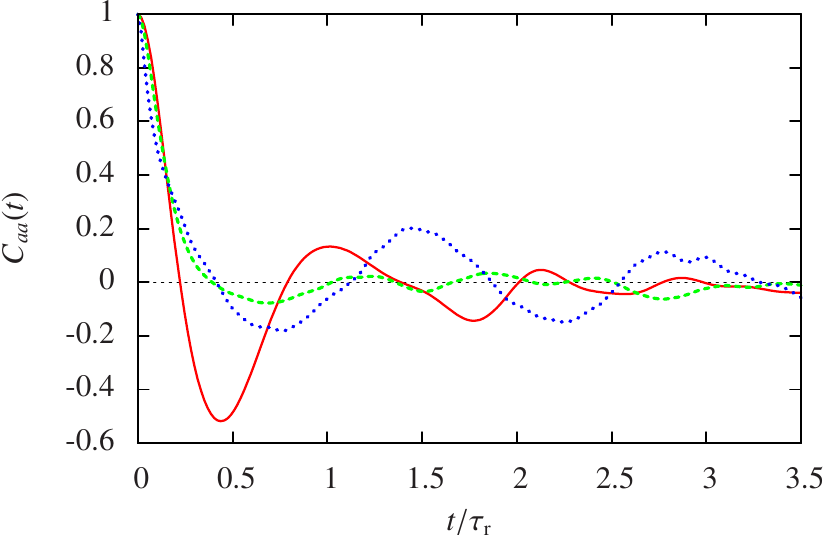}
\caption{\label{fig:corrTTurbPic}(Color online) Time autocorrelation 
functions for different Damk\" ohler numbers $\mathrm{Da}$. Estimates 
of $C_{aa}(t)$ were obtained for $\mathrm{Da}=0.29$ (dotted blue line), 
$\mathrm{Da}=0.92$ (dashed green line), and $\mathrm{Da}=4.7$ 
(solid red line). The diffusion to reaction time scale ratio 
$K_{\mathrm d}$ was set to $K_{\mathrm d}=3.2\times10^3$ in all cases.}
\end{figure}

The time autocorrelations $C_{ss}(|t-t'|)\equiv \mathop{C_{ss}}(0, t, t')$ 
were initially obtained from a time average over time 
$T\approx 11\tau_{\mathrm r}$, taken at a fixed point inside the simulation 
domain, once the total density had reached a steady value. By examining the 
initial estimates, we realized that the chosen averaging time was too short 
to give satisfactory results. Therefore, we picked three points from 
different regions of the parameter space and improved our estimates of 
$C_{ss}(|t-t'|)$---for those three particular choices of control 
parameters---by taking an additional average over 15 realizations of the 
model. The final results for $C_{ss}(|t-t'|)$ are shown in 
Fig.~\ref{fig:corrTTurbPic}. The estimated time autocorrelations are not as 
well behaved as their corresponding spatial part, but we are still confident 
that the results in Fig.~\ref{fig:corrTTurbPic} are sufficiently accurate to 
correctly predict the gross features of the time autocorrelations. Namely, 
$C_{ss}(|t-t'|)$ display damped oscillations, except for values of 
$\mathrm{Da}$ around $\mathrm{Da}\approx1$. In the low $\mathrm{Da}$ regime, 
the oscillations in $C_{ss}(|t-t'|)$ arise from the collective oscillations 
in relative species abundance, whereas in the high $\mathrm{Da}$ regime, 
the oscillations result from the (unstable) rotating spiral patterns. For 
intermediate values of $\mathrm{Da}$ around $\mathrm{Da}\approx 1$, the time 
autocorrelations quickly decay towards zero without any clear signs of 
(damped) oscillations. This result justifies the use of the spirals' rotation
period $\mathop{T_0}$ for the definition of the reaction time scale 
$\tau_{\mathrm r}$ because the reactions appear to be approximately in 
balance with turbulent advection when $\mathrm{Da}\approx 1$ 
(i.e., when $T_0\approx\tau_{\mathrm f}$).

\subsection{\label{ssec:ecosystemStability} Global attractors 
and ecosystem stability}

The qualitative changes in the system's spatiotemporal dynamics with respect
to the control parameters seem very pronounced, suggesting that the model's 
parameter space can be divided into different dynamical regimes. To develop 
our idea further, we follow the approach of \mbox{Rulands} et 
al.~\cite{Rulands2013} and analyze the attractors of the global 
(space-averaged) dynamics. Adopting the terminology of 
Ref.~\cite{Rulands2013}, we call the attractors of the space-averaged 
dynamics \emph{global attractors}. To avoid confusion, we note that the same 
phrase is also used in a similar context in mathematical literature but its 
specific meaning there is different. Due to the inherent presence of 
flow-induced statistical fluctuations, the global attractors correspond to 
maxima of the probability density to find the system in a specific global 
state $(\mathop{\overline a}, \mathop{\overline b}, \mathop{\overline c})$ 
rather than to isolated orbits or points, for example. In this framework, 
the transitions between different dynamical regimes are to be interpreted 
as bifurcations of the global dynamics and should not be confused with 
nonequilibrium phase transitions. Moreover, the observed qualitative 
changes in the dynamics arise essentially from finite-size effects due 
to variations of the physical length scales compared to the domain size. 
Evidently, the global attractors do not give any information regarding 
the small-scale variability of the subpopulation densities. However, 
the dimensional reduction of the problem considerably simplifies 
the search for bifurcations in the parameter space, while at the same time 
it still gives valuable insight into the nature of biological interactions 
on the largest scales. 

\setcounter{figure}{8}
\begin{figure*}[ht]
\includegraphics{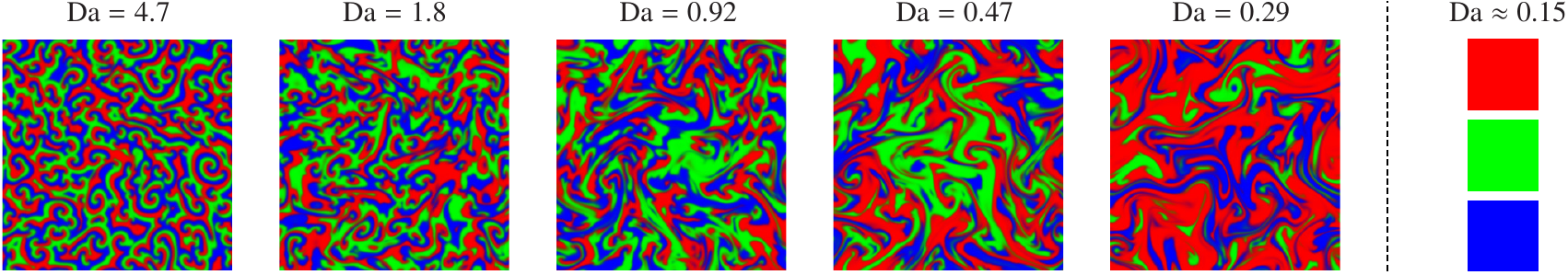}
\caption{\label{fig:extPic}(Color online) Approach to the absorbing state 
with a decreasing Damk\" ohler number $\mathrm{Da}$ at a fixed 
diffusion to reaction time scale ratio $K_{\mathrm d}=8.1\times10^3$. 
Each subpopulation density is represented by its own color 
channel (gray tone) as described in Fig.~\ref{fig:RGBplots}. As 
illustrated above, a relative increase of turbulent advection enhances the 
dispersal of species and increases the average size of spatial patterns, 
until the biodiversity is lost. For $K_{\mathrm d}=8.1\times10^3$, we were 
able to observe the first extinction events around $\mathrm{Da}\approx 0.15$.}
\end{figure*}
\setcounter{figure}{7}

\begin{figure}[ht]
\includegraphics{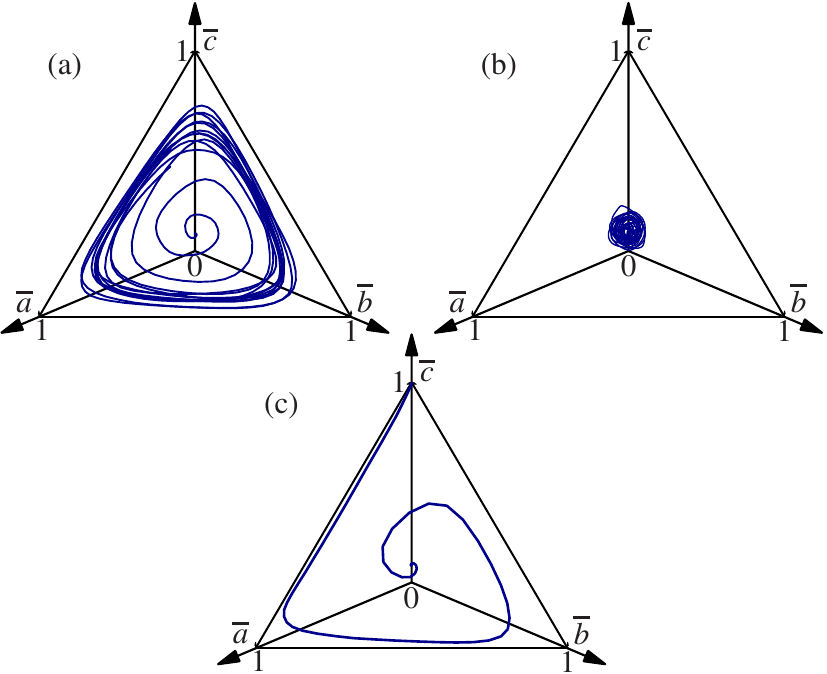}
\caption{\label{fig:globalAttractors}(Color online) Global phase portraits 
of the dynamics for different Damk\" ohler numbers $\mathrm{Da}$ and 
ratios of the diffusion to reaction time scale $K_{\mathrm d}$. 
The trajectories of the space-averaged solutions are shown 
for $\mathrm{Da}=0.21$, $K_{\mathrm d}=8.1\times10^3$ (a); 
$\mathrm{Da}=0.71$, $K_{\mathrm d}=8.1\times10^3$ (b); 
and $\mathrm{Da}=0.29$, $K_{\mathrm d}=5.4\times10^2$ (c).}
\end{figure}

Figure \ref{fig:globalAttractors} shows three global phase portraits of the 
solutions, corresponding to three different types of global attractors 
observed in our simulations. The attractors can be readily identified as 
a ``limit cycle'' 
(Fig.~\hyperref[fig:globalAttractors]{\ref*{fig:globalAttractors}(a)}), 
a ``fixed point'' 
(Fig.~\hyperref[fig:globalAttractors]{\ref*{fig:globalAttractors}(b)}), 
and a heteroclinic orbit 
(Fig.~\hyperref[fig:globalAttractors]{\ref*{fig:globalAttractors}(c)}). 
The solutions starting in the vicinity of the reactive fixed point of 
Eq.~\eqref{eq:rps} converge (in a statistical sense) to a limit cycle or to 
the fixed point attractor if turbulent mixing is not too strong so that 
biodiversity can be maintained on the time scales of interest. It is worth 
emphasizing that the fixed point global attractor does not coincide with the 
reactive fixed point of the rate equations. Instead, it is shifted along the 
symmetry axis 
$\mathop{\overline a}=\mathop{\overline b}=\mathop{\overline c}$ towards a 
higher density $\mathop{\overline\rho}\approx 0.9$ as compared to the 
well-mixed limit where $\rho=3/4$ in the reactive fixed point. The limit 
cycle global attractors correspond to oscillations in relative species 
abundance observed for low $\mathrm{Da}$. In the context of dynamical 
systems theory, the collective oscillations are quite a remarkable property,
considering the fact that the well-mixed system \eqref{eq:rps} is 
characterized by heteroclinic orbits rather than by limit cycles. 
Neglecting the mathematical details regarding the true asymptotic nature of 
solutions corresponding to heteroclinic orbits, we may say that the 
heteroclinic orbits correspond to extinctions of all but one species. In 
the spatially extended model \eqref{eq:rpsTurb}, the system tends to get 
trapped into one of the absorbing states when the average size of spatial 
patterns approaches the domain size (Fig.~\ref{fig:extPic}). The size of 
spatial patterns is, in turn, controlled by $\mathrm{Da}$ and 
$K_{\mathrm d}$. Therefore, species coexistence depends essentially 
on the choice of $\mathrm{Da}$ and $K_{\mathrm d}$.

\setcounter{figure}{9}

\begin{figure}[ht]
\includegraphics{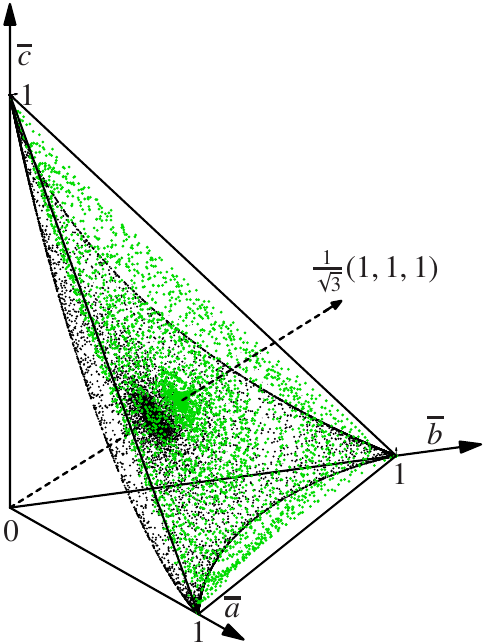}
\caption{\label{fig:invManifoldAvg}(Color online) The invariant manifold of 
Eq.~\eqref{eq:rps} (black dots) compared to the solutions of 
Eq.~\eqref{eq:rpsTurb} projected onto the global phase space 
(green/gray dots). The dashed arrow denotes the unit vector along the 
symmetry axis.}
\end{figure}

A common feature shared by all global attractors of the spatially extended 
model is their confinement to a quasi-2D geometry within the 
three-dimensional global phase space. In Fig.~\ref{fig:invManifoldAvg}, we 
show the projections of long-time solutions for various $\mathrm{Da}$ and 
$K_{\mathrm d}$ onto the global phase space and compare the obtained result 
with the invariant manifold of Eq.~\eqref{eq:rps}. It is shown that the 
global dynamics can be effectively reduced to a quasi-2D surface within the 
global phase space which does not correspond to the invariant manifold of 
Eq.~\eqref{eq:rps}. The quasi-2D surface spanned by the solutions of 
Eq.~\eqref{eq:rpsTurb} has a very mild curvature at the intersection with 
the symmetry axis around $\mathop{\overline\rho}\approx0.9$ and becomes 
slightly more curved close to the boundaries of the phase space where it 
touches the single-species equilibrium points. For most values of 
$\mathrm{Da}$ and $K_{\mathrm d}$ considered in our simulations, even the 
solutions that correspond to transitions into absorbing states lie on this 
surface rather than on the invariant manifold of the well-mixed system. A 
significant departure of the phase portraits from the global surface of 
solutions was observed only for $K_{\mathrm d}\sim 10$, and it should be 
probably required that $K_{\mathrm d}\sim 1$  in order to completely neglect
the spatial degrees of freedom.

\begin{figure}[ht]
\includegraphics{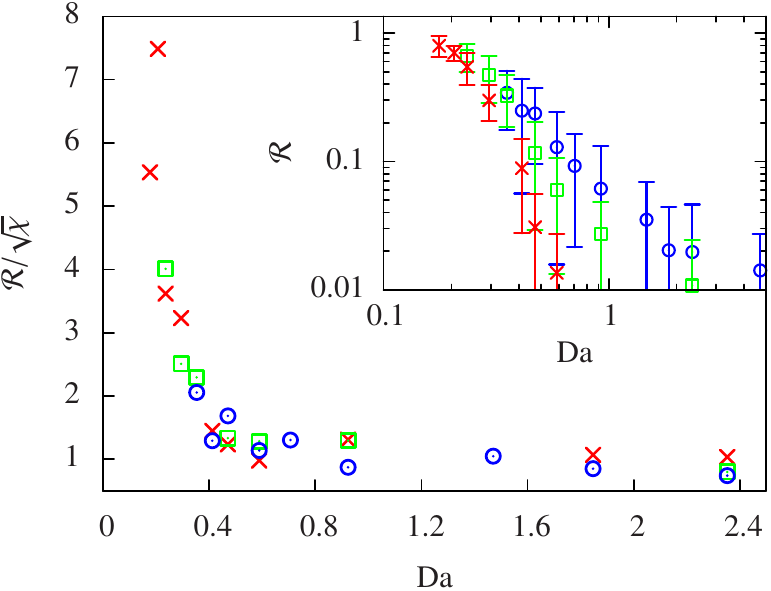}
\caption{\label{fig:collectiveOscillations}(Color online) Dependence of 
the rescaled radius ${\mathcal R}/\sqrt\chi$ on the Damk\" ohler 
number $\mathrm{Da}$ for different diffusion to reaction time scale 
ratios $K_{\mathrm d}$. Estimates of ${\mathcal R}/\sqrt\chi$ 
are shown for $K_{\mathrm d}=8.1\times10^3$ (red crosses), 
$K_{\mathrm d}=1.8\times10^3$ (green squares), and 
$K_{\mathrm d}=5.4\times10^2$ (blue circles). The inset shows the 
measurements of $\mathcal R$ with the error bars representing the 
magnitude of statistical fluctuations $\sqrt{\chi}$.}
\end{figure}

Let us now try to analyze the transition into the collective oscillations 
regime in more detail. Due to the inherent presence of noise in the system, 
it is necessary to use a statistical approach to distinguish between the 
fixed point and limit cycle global attractor. In order to distinguish 
a limit cycle from small fluctuations around the symmetry axis which 
correspond to the fixed point attractor, the global trajectories have to be 
well separated from the symmetry axis 
$\mathop{\overline a}=\mathop{\overline b}=\mathop{\overline c}$. To 
introduce a measure for the mean separation of trajectories from the symmetry
axis, we can make use of the Lyapunov function of the global concentrations 
$\mathcal{L}\equiv 
(\mathop{\overline a}\mathop{\overline b}\mathop{\overline c})
/(\mathop{\overline a}+\mathop{\overline b}+\mathop{\overline c})^3$, which 
can be regarded as a radial coordinate, measuring the distance of a point 
from the boundaries of the global phase space 
\cite{Hofbauer1998, Rulands2013}. Using the Lyapunov function, an effective 
radius of a limit cycle as measured from the symmetry axis can be defined as
${\mathcal R}\equiv 1 - {\mathcal L}/\mathcal{L}_{\max}$, where 
$\mathcal L_{\max}=1/27$. Hence, to distinguish a fixed point 
from a limit cycle in a statistical sense, we should at 
least require that $\mathcal{R}/\sqrt{\chi} \gtrsim 1$ 
in the collective oscillations regime, where 
$\chi = \langle{\mathcal R^2}\rangle - \langle{\mathcal R}\rangle^2$. The 
dependence of the rescaled radius $\mathcal{R}/\sqrt{\chi}$ on $\mathrm{Da}$ 
is shown in Fig.~\ref{fig:collectiveOscillations}. The abrupt jump around 
$\mathrm{Da}_c\approx 0.4$ reflects the underlying bifurcation of the global 
dynamics. We do not rule out a possible weak dependence of the transition 
also on $K_{\mathrm d}$, but the available simulation data are insufficient 
to clearly confirm or neglect a possible dependence on $K_{\mathrm d}$.

\begin{figure}[ht]
\includegraphics{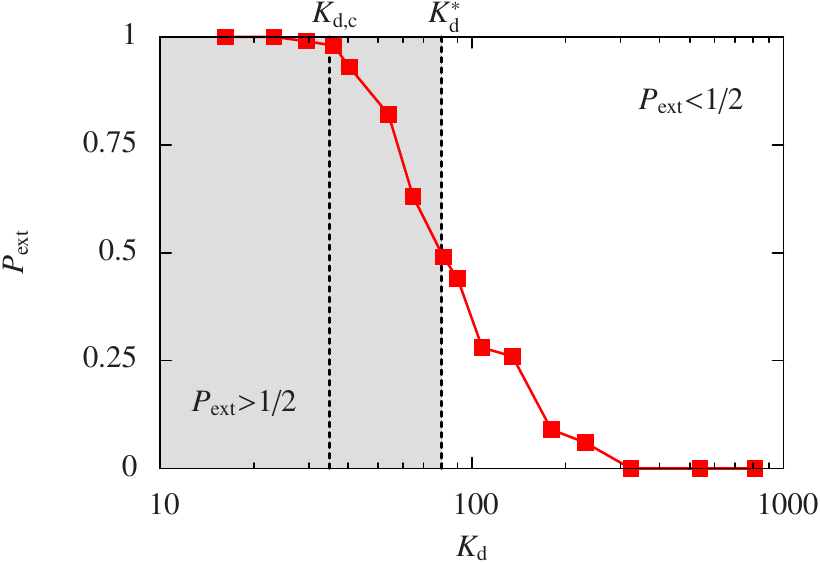}
\caption{\label{fig:deathProb}(Color online) Extinction probability 
$P_{\mathrm{ext}}$ calculated in the absence of fluid flow as a function 
of the diffusion to reaction time scale ratio $K_{\mathrm d}$. The values 
$K_{\mathrm{d,c}}$ and $K_{\mathrm d}^*$ on the graph denote the points 
above which the extinction probability drops below 
$P_{\mathrm{ext}}\approx 1$ and $P_{\mathrm{ext}}\approx 1/2$, respectively.}
\end{figure}

To quantitatively describe the transitions into absorbing states, one 
can consider the extinction probability $P_{\mathrm{ext}}$ that two species 
have gone extinct after time $T$ \cite{Reichenbach2007}. For the sake of 
simplicity, and for consistency with previous works 
\cite{Reichenbach2007, Rulands2013} performed for the discretized version of 
model \eqref{eq:rpsTurb} in the absence of fluid flow, let us first discuss 
the limiting case $\mathrm{Da}\!\to\!\infty$. To begin with, the meaning of 
the extinction probability in deterministic reaction-diffusion models 
requires some special attention. The randomness in discrete models originates
from the stochastic nature of biological interactions, whereas in models 
described by partial differential equations (PDEs) the ``randomness'' can be 
achieved by considering various initial conditions. These two formulations 
might at a first glance appear as completely unrelated, however, our approach 
illustrates that the results of the PDE model are equivalent to the ones 
produced from discrete lattice simulations with a large number of particles 
\cite{Reichenbach2007, Reichenbach2008}, provided that the initial conditions 
match the solutions of the discrete lattice model at early stages of the 
system's time evolution. In our case, this means that the initial conditions 
for the PDE model should be generated as random, $\delta$-correlated-in-space
perturbations around the reactive fixed point of Eq.~\eqref{eq:rps}. In 
Fig.~\ref{fig:deathProb}, we show the dependence of $P_{\mathrm{ext}}$ on 
$K_{\mathrm d}$ in the limit $\mathrm{Da}\!\to\!\infty$. Since the spatial 
patterns of the reaction-diffusion system remain unchanged for all times 
after an initial transient of a typical duration 
$T\!\sim\!10\tau_{\mathrm r}$, we consider a waiting time 
$T\approx9.8\tau_{\mathrm r}$, and each estimate for $P_{\mathrm{ext}}$ 
is obtained from an average over 100 initial conditions. The critical 
value $K_{\mathrm{d,c}} = 35\pm5$, below which species coexistence 
becomes almost impossible, is in good agreement with the result from 
\mbox{Reichenbach} et al.~\cite{Reichenbach2007} obtained from 
discrete lattice simulations (written in terms of $K_{\mathrm d}$, the 
estimate from Ref.~\cite{Reichenbach2007} 
reads $K_{\mathrm{d,c}} = 36\pm 4$).

A large number of simulations are required for an accurate estimate of the 
threshold $K_{\mathrm{d,c}}$ because the survivals of all three 
competitors are statistically very rare near $K_{\mathrm{d,c}}$. If one 
only requires rough estimates of the transition points into the absorbing 
states, a better alternative is to consider a threshold 
$K_{\mathrm d}^*$ for which $\mathop{P_{\mathrm{ext}}}\approx 1/2$ 
(see Fig.~\ref{fig:deathProb}). Indeed, rough estimates 
of $K_{\mathrm d}^*$ can be obtained even if only a single simulation run 
is performed for each choice of the control parameters because the point 
$K_{\mathrm d}^*$ should be in any case bounded from above by the values of 
$K_{\mathrm d}$ for which $P_{\mathrm{ext}}\approx 0$ and from below by 
those values for which $P_{\mathrm{ext}}\approx 1$. Therefore, in order to 
spend our computational resources wisely, we have not explicitly considered 
the extinction probabilities for the more general case of a nonzero fluid 
velocity since it is already possible to make qualitative conclusions 
regarding ecosystem stability from a very limited number of simulation runs, 
performed for various choices of $\mathrm{Da}$ and $K_{\mathrm d}$. For the 
reasons explained above, we use the threshold 
$\mathop{P_{\mathrm{ext}}}\approx 1/2$ as a more formal definition of 
the boundary between the species coexistence region and the extinction 
region. It should be emphasized that in the presence of random fluctuations, 
such as the ones arising from a turbulent flow, the ecosystem can never 
remain stable in the strict limit $T\to\infty$ because there always exists 
a possibility that the random disturbances will drive the system into one of 
its absorbing states. The global attractors from the species coexistence 
region are therefore in a strict mathematical sense only long-lived 
transients. However, for sufficiently large $K_{\mathrm d}$ and 
$\mathrm{Da}$ the extinction probability becomes very small on 
the biological time scales of interest, and the states 
corresponding to these parameters may be for practical 
purposes regarded as states of species coexistence \cite{Hastings2004}.

\begin{figure}[ht]
\includegraphics{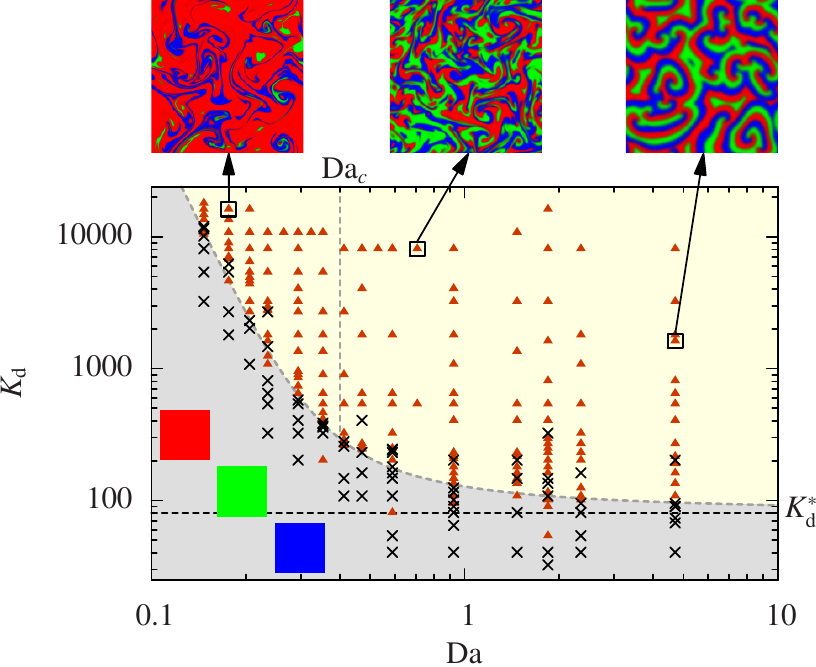}
\caption{\label{fig:phaseDiag}(Color online) Nonequilibrium phase diagram 
of the spatiotemporal dynamics. The black crosses show the simulation runs 
where biodiversity was lost before $T\approx23\tau_{\mathrm r}$, and the 
brown triangles show the survivals of all three species. The (medium) gray 
shading indicates the region of extinction where transitions into absorbing 
states become more probable than long-lived states of biodiversity. The 
horizontal and vertical dashed lines show the thresholds $K_{\mathrm d}^*$ 
and $\mathrm{Da}_c$, respectively (see text for further details). Above the 
phase diagram we show three long-time snapshots of the solutions from 
different parts of the species coexistence region.}
\end{figure}

To determine the relative extent of the species coexistence region in the 
parameter space, we performed several simulations for various 
$\mathrm{Da}$ and $K_{\mathrm d}$. In each of these simulations we 
integrated Eq.~\eqref{eq:rpsTurb} for at least $T\approx 23\tau_{\mathrm r}$.
Figure \ref{fig:phaseDiag} finally summarizes the main global dynamical 
features of our spatially extended ecosystem with a nonequilibrium phase 
diagram. Based on how the species extinction events and their survivals 
are distributed across the parameter space, it is possible to sketch a 
rough dependence of the $P_{\mathrm{ext}}\approx 1/2$ extinction probability 
threshold on $\mathrm{Da}$ and $K_{\mathrm d}$. Below the transition 
into the collective oscillations regime around $\mathrm{Da}_c$, the turbulent
flow becomes increasingly capable of synchronizing the local subpopulation 
density oscillations among distant parts of the domain, which in turn greatly
reduces the extent of the species coexistence region with respect to 
$K_{\mathrm d}$. For high $\mathrm{Da}$, our results seem to be consistent 
with the limiting value $K_{\mathrm d}^*$, estimated from the simulations 
performed in the absence of fluid flow.

\section{\label{sec:conclusions} Conclusions}

We studied the population dynamics of three cyclically
competing species in a two-dimensional turbulent flow forced at large scales.
The presented results of our numerical simulations give new insight
into how turbulent transport affects ecosystem structure in biological
communities without a clear competition hierarchy. More specifically, we
performed simulations over a broad range of relative advection and
diffusion strengths compared to the biological reactions and studied 
how different choices of ecosystem parameters affect the system's
spatiotemporal dynamics and species' biodiversity. For short reaction time 
scales $\tau_{\mathrm r}$ compared to the characteristic flow time 
scale $\tau_{\mathrm f}$, corresponding to large Damk\" ohler numbers 
$\mathrm{Da} = \tau_{\mathrm f}/\tau_{\mathrm r}$, the subpopulations 
self-organize into rotating spiral waves. This phenomenon is consistent 
with previous studies of cyclic competitions in reaction-diffusion
systems, with the exception that the spiral patterns become unstable when 
subjected to perturbations induced by a turbulent flow. For Damk\" ohler 
numbers around $\mathrm{Da}\approx 1$, with $\tau_{\mathrm r}$ defined as 
the spirals' rotation period measured in the absence of fluid flow, the 
reaction-diffusion dynamics and turbulent advection are found to be in an 
approximate dynamic balance. This is most clearly seen by inspecting the 
species (Eulerian) time autocorrelation functions, which do not show 
any clear signs of periodic oscillations for $\mathrm{Da}\approx 1$. 
When the Damk\" ohler number is decreased even further, a sharp transition 
to a state with collective oscillations in relative species abundance is 
observed at a certain threshold value of $\mathrm{Da}$. The observed 
phenomenon suggests that turbulence might play an important role in the 
structuring of marine phytoplankton communities, which are typically 
composed of only a few dominant species while the remaining ones 
represent only a small fraction of the total biomass \cite{Pommier2007}. 
This type of interpretations are also supported by some recent 
numerical experiments presented in Ref.~\cite{Levy2014}.

To further investigate the transitions between the qualitatively different 
dynamical regimes observed in our simulations, we studied the attractors of 
the global (space-averaged) dynamics and identified three different 
types attractors, corresponding to maxima of the probability density to 
find the system in a specific global (space-averaged) state. The three 
different types of global attractors have been identified as a fixed point, 
limit cycles, and heteroclinic orbits. The transitions between these attractors 
should be interpreted as bifurcations of the global dynamics. The fixed point 
attractor corresponds to states of species coexistence observed for Damk\" ohler 
numbers of order unity and above, whereas the limit cycles correspond to 
collective oscillations observed for low $\mathrm{Da}$. The heteroclinic orbits 
indicate transitions to homogeneous states, where the system's biodiversity is 
lost. The probability that two of the species will go extinct depends 
essentially on the choice of ecosystem parameters. Outside the regime of 
collective oscillations, the extinction probability depends most strongly on 
the relative strength of diffusion compared to the reactions. However, as the 
Damk\" ohler number drops below a certain threshold, turbulence becomes 
increasingly capable of synchronizing oscillations among distant parts of the 
domain, which greatly reduces the acceptable range of relative diffusion 
strengths that still allow for species coexistence.

In all of our simulations, we used the same type of turbulent flow. Further 
studies could investigate the effects of different types of flows on 
cyclic interactions, and identify which system properties are more general 
and which ones depend on the details of the turbulence model. In particular, 
it would be interesting to consider the dependence of solutions on the 
correlation length of the flow, since it would seem reasonable to expect 
that the turbulent flow becomes incapable of producing collective 
oscillations in relative species abundance when the correlation length 
is much smaller than the domain size. To shed some light on these 
speculations, we performed a couple of trial simulations at 
$\ell_{\mathrm f}=L/10$. There, we were still able to observe the collective 
oscillations but their amplitude for the same choice of $\mathrm{Da}$ 
was somewhat smaller.

Finally, it should also be noted that, even though the main motivation for 
this work comes from the field of theoretical population biology, the 
presented results are also relevant in other fields where RDA systems of 
a similar type as the one studied here can be found. In particular, 
pattern-forming reactions in fluid flows have also been realized in 
experiments with the chemical Belousov-Zhabotinsky
reaction~\cite{Nugent2004, Paoletti2006, Kameke2013}.

See Ref.~\footnote{The simulation source code used in this study is 
publicly available at \url{https://github.com/dgroselj/DRASLA}} 
for information regarding the project's source code. 

\begin{acknowledgments}

The research leading to these results has received funding from the
European Research Council under the European Union's Seventh Framework
Programme (FP7/2007-013)/ERC Grant Agreement No.~277870.
This research was also supported by the German Excellence Initiative 
via the program ``Nanosystems Initiative Munich'' and the Deutsche 
Forschungsgemeinschaft via Contract No.~FR 850/9-1.
We thank V.~\mbox{Bratanov}, K.~\mbox{Reuter}, 
T.~\mbox{G\" orler}, and E.~\mbox{Sonnendr\" ucker} for useful 
discussions. We also thank M.~\mbox{Perc} for comments on the 
manuscript.

\end{acknowledgments}

\bibliography{\jobname}

\end{document}